\documentclass{INTERSPEECH2023}
\usepackage{multirow}
\usepackage{setspace}

\interspeechcameraready

\title{UNISOUND System for VoxCeleb Speaker Recognition Challenge 2023}
\name{Yu Zheng$^1$$^,$$^*$, Yajun Zhang$^1$$^,$$^*$, Chuanying Niu$^1$, Yibin Zhan$^1$, Yanhua Long$^2$, Dongxing Xu$^1$}
\address{
  $^1$Unisound AI Technology Co., Ltd., Beijing, China\\
  $^2$SHNU-Unisound Joint Laboratory of Natural Human-Computer Interaction, Shanghai Normal University, Shanghai, China}
\email{zhengyu@unisound.com, xudongxing@unisound.com}

\begin{document}

\maketitle
\begingroup\renewcommand\thefootnote{*}
\footnotetext{These authors share equal contribution to this work. Yanhua Long is also with the Key Innovation Group of Digital Humanities Resource and Research, Shanghai Normal University.}

\begin{abstract}
This report describes the UNISOUND submission for Track1 and Track2 of VoxCeleb Speaker Recognition Challenge 2023 (VoxSRC 2023). We submit the same system on Track 1 and Track 2, which is trained with only VoxCeleb2-dev. Large-scale ResNet and RepVGG architectures are developed for the challenge. We propose a consistency-aware score calibration method, which leverages the stability of audio voiceprints in similarity score by a Consistency Measure Factor (CMF). CMF brings a huge performance boost in this challenge. Our final system is a fusion of six models and achieves the first place in Track 1 and second place in Track 2 of VoxSRC 2023. The minDCF of our submission is 0.0855 and the EER is 1.5880\%.

\end{abstract}

\section{Dataset}
We only used VoxCeleb2-dev \cite{chung2018voxceleb2} as training data without any extra data for both Track 1 and Track 2, which contained 5,994 speakers with 1,092,009 utterances.

\subsection{Data augmentation}
 Firstly, we adopted a 3-fold speed augmentation to generate extra twice speakers. Each speech segment was perturbed by 0.9 and 1.1 factor based on the SoX speed function. Secondly, we used RIRs \cite{ko2017study} and MUSAN \cite{snyder2015musan} to create extra four copies of the training utterances and the data augmentation process was based on the Kaldi \cite{povey2011kaldi, snyder2018x} VoxCeleb recipe. Finally we obtained 1,7982 speakers with 16,380,135 utterances.

\subsection{Features}
We extracted 80-dimensional log Mel filter bank with energy using Kaldi toolkit. 
The window size was 25 ms with a 10 ms frame shift.  No voice activity detection (VAD) was applied. Chunks of features were mean-normalized before fed into the network.

\section{System}

\subsection{Architectures}

We used a sequence of large-scale ResNet \cite{he2016deep} and RepVGG \cite{ding2021repvgg,zhao2021speakin} architectures as backbones of our systems.  Multi-query multi-head attention (MQMHA) \cite{mqmha_intertopk} pooling layer was attached after backbone in every system. A fully connected feed-forward layer with 512 dimensions is added after the pooling layer, and the output of this layer is the embedding of the input audio. The last connection of the network is an AM-Softmax \cite{wang2018additive, wang2018cosface} or AAM-Softmax \cite{deng2019arcface}.

\subsubsection{ResNet}

ResNet structure was widely used in speaker recognition systems. In addition to the classic ResNet structure, we increased the number of ResNet layers to 242, 314, and 518. Like the classic ResNet, after a 3$\times$3 convolution layer as the first layer, the network structure consists of four convolution blocks. Each block was stacked with bottlenecks \cite{he2016deep}. The strides of blocks was (1, 2, 2, 2). Number of channels of blocks was (64, 128, 256, 512). The number of bottleneck in every block of  ResNet242, 314, 518 architecture were shown in Table~\ref{tab:ResNets}.

Including the MQMHA and embedding layers, the largest model, ResNet518 has a total of 227.46M parameters.

\begin{table}[]
\caption{ResNets architecture}
\label{tab:ResNets}
\centering
\begin{tabular}{ccc}
\toprule
\textbf{Name} & \textbf{Layers of each block} & \textbf{Params(M)} \\ 
\midrule
ResNet101     & 3, 4, 23, 3                      & 42.49              \\
ResNet152     & 3, 8, 36, 3                      & 58.13              \\
ResNet242     & 3, 10, 64, 3                     & 89.98              \\
ResNet314     & 3, 16, 82, 3                     & 111.77             \\
ResNet518     & 6, 32, 128, 6                    & 181.24             \\ 
\bottomrule
\end{tabular}
\end{table}

\subsubsection{RepVGG}

 RepVGG as one of the re-parameterized models, shows competitive performance in speaker recognition \cite{zhao2021speakin}. We select RepVGG-B1 with 64 base channels as backbone of a subsystem in this challenge. 

\subsubsection{Pooling layer}
Multi-query multi-head attention (MQMHA) was used in each system. And the number of query was set to 4 while the number of head was 16. 

Different from the classic pooling layer, we used MQMHA to calculate only the standard derivation along the time dimension in all ResNet systems. However, in 
the RepVGG system, both the mean and the standard derivation were calculated. 
Morevoer, We found that using only standard deviation in the pooling layer of the ResNet systems lead to better performance, but it made performance of RepVGG sytems much worse.

\subsubsection{Loss function}
We used Large-Margin Softmax in each system. AM-Softmax and AAM-Softmax were used in different stages of training. The Sub-Center method \cite{deng2020sub} was introduced, and the number of center was set to 3. We also used the Inter-TopK \cite{mqmha_intertopk} penalty to pay further attention to difficult samples.

\begin{table*}[t]
\centering
\caption{Results on Development Sets}
\label{tab:tabel_results}
\setlength{\tabcolsep}{1.5mm}{
\begin{tabular}{llcccccccc}
\toprule
\multirow{2}{*}{\textbf{Index}} & \multicolumn{1}{c}{\multirow{2}{*}{\textbf{System}}} & \multicolumn{2}{c}{\textbf{VoxCeleb1-O}}    & \multicolumn{2}{c}{\textbf{VoxCeleb1-E}}    & \multicolumn{2}{c}{\textbf{VoxCeleb1-H}}    & \multicolumn{2}{c}{\textbf{VoxSRC23-val}}   \\
\cmidrule(lr){3-10}
                                & \multicolumn{1}{c}{}                                 & \textbf{EER(\%)}     & \textbf{DCF$_{0.01}$}   & \textbf{EER(\%)}     & \textbf{DCF$_{0.01}$}   & \textbf{EER(\%)}     & \textbf{DCF$_{0.01}$}   & \textbf{EER(\%)}     & \textbf{DCF$_{0.05}$}   \\ \midrule 
\textbf{S1}                     & ResNet34                                             & 0.5037             & 0.0546              & 0.7123             & 0.0597              & 1.1246             & 0.0990              & 2.4956             & 0.1394               \\
\textbf{S2}                     & ResNet101                                            & 0.4136             & \textbf{0.0293}     & 0.6088             & 0.0450              & 0.9722             & 0.0755              & 2.1642             & 0.1210               \\
\textbf{S3}                     & ResNet152                                            & 0.4401             & 0.0313              & 0.6047             & 0.0454              & 0.9313             & 0.0702              & 2.0121             & 0.1143               \\
\textbf{S4}                     & ResNet242                                            & 0.4348             & 0.0370              & \textbf{0.5789}    & 0.0413              & \textbf{0.9042}    & 0.0666              & 2.0082             & 0.1188               \\
\textbf{S5}                     & ResNet314                                            & 0.4454             & 0.0361              & 0.6339             & 0.0465              & 0.9701             & 0.0760              & 1.9809             & \textbf{0.1072}               \\
\textbf{S6}                     & ResNet518                                            & \textbf{0.3712}    & 0.0299              & 0.5851             & \textbf{0.0391}     & 0.9093             & \textbf{0.0647}     & \textbf{1.8678}    & 0.1074      \\
\textbf{S7}                     & RepVGG-B1                                            & 0.4348             & 0.0368              & 0.6425             & 0.0549              & 1.0019             & 0.0793              & 2.2266             & 0.1290               \\ \midrule
\multicolumn{2}{l}{\textbf{Fusion\quad S2$\sim$S7}}                                              & \textbf{0.3659}    & \textbf{0.0241}     & \textbf{0.5651}    & \textbf{0.0364}     & \textbf{0.8662}    & \textbf{0.0587}     & \textbf{1.8327}    & \textbf{0.1048}  
\\ \bottomrule
\end{tabular}}
\end{table*}

\subsection{Backend}

\subsubsection{CMF score calibration}
Inspired by segment scoring, which samples ten 3-second temporal crops from each test segment and uses the mean of distances between the every possible pair of crops as the final score\cite{Chung_2018}. We assume a simplest case to derive this procedure. Suppose $\bm{y}$ is a embedding of audio A, and 
$(\bm{x_{1}}, \bm{x_{2}}, ..., \bm{x_{N}})$ are the embeddings of $N$ segments from audio B which cut into $N$ crops. The segment score of audio A and audio B is:

\begin{equation}
s_{AB}=\frac{1}{N} \sum_{i=1}^{N}\cos(x_{i}, y)
= \frac{y}{N\cdot||y||}\sum_{i=1}^{N}\frac{x_{i}}{||x_i||}
\end{equation}

\begin{equation}
c=\sum_{i=1}^{N}\frac{x_{i}}{||x_i||}
\end{equation}

\begin{equation}
s_{AB}=\frac{y}{N \cdot||y||}\cdot c=\frac{1}{N}||c||\cdot cos(y,c)
\label{eq3}
\end{equation}

\begin{equation}
CMF_{B}=\frac{1}{N}||c||
\end{equation}

We define $\frac{1}{N}||c||\in(0,1)$ as Consistency Measure Factor (CMF). CMF reflects the degree of consistency or dispersion of the embeddings $(x_{1}, x_{2}, ..., x_{N})$. As Figure \ref{CMF} shows, larger value of CMF indicates that the distribution of vectors is more concentrated. To some extent, CMF reflects the stability of audio voiceprint.

\begin{figure}[t]
  \centering
  \includegraphics[width=0.44\textwidth,height=0.2\textwidth]{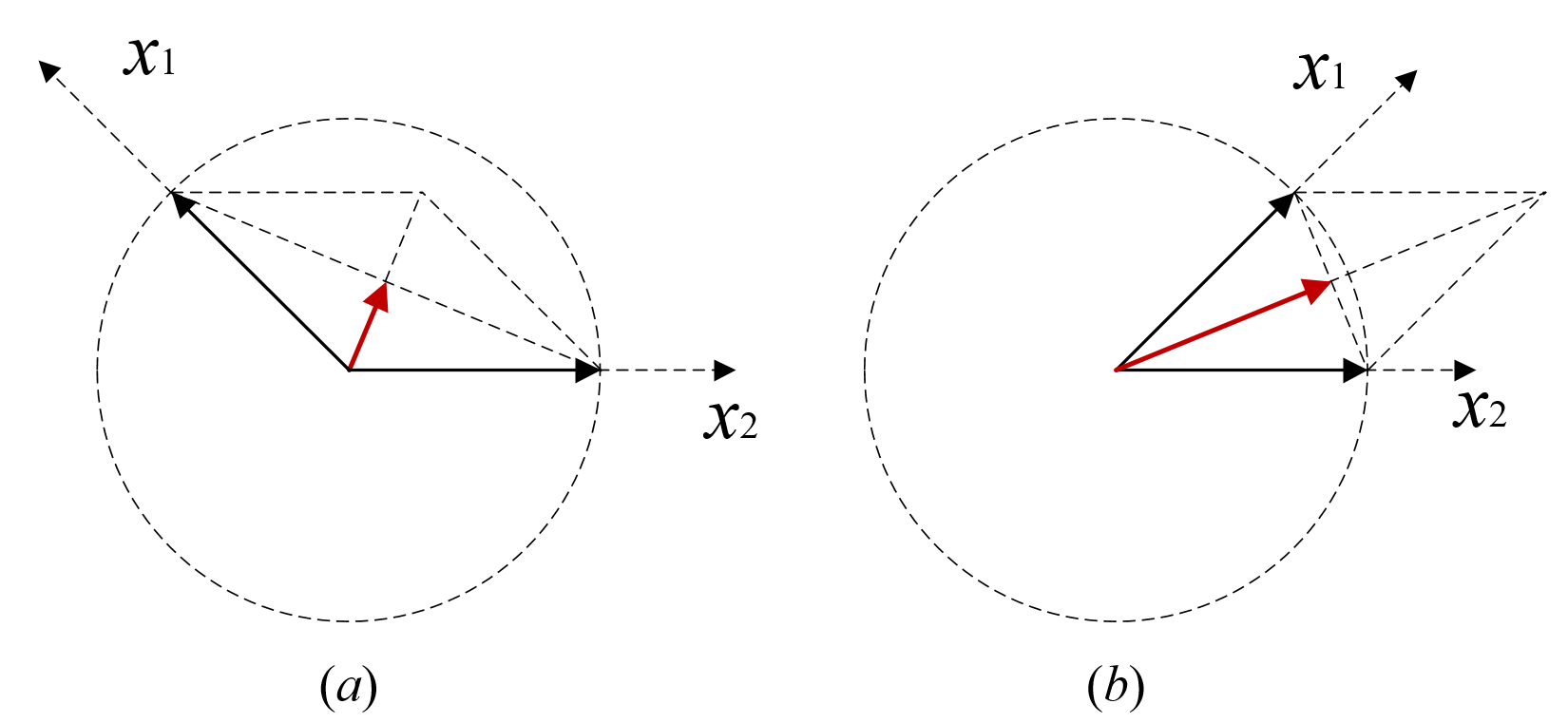}
  \caption{\textbf{The connection between CMF and segment embeddings}. Suppose $N=2$ and the dimension of $x_{i}$ is 2, and the red arrow represents  $\frac{1}{N}c$. (a) When the angle between $x_{1}$ and $x_{2}$ is large, the modulus length of $\frac{1}{N}c$ is smaller; (b) When the cosine distance between $x_{1}$ and $x_{2}$ is large, the modulus length of $\frac{1}{N}c$ is large. }
  \label{CMF}
\end{figure}

There is a problem with segment scoring, when the segment length is shorter, it may not be friendly to judge the similarity, but it may increase the discrimination of CMF. To fix this problem, we only use CMF as a scale to calibrate score as Eq.(\ref{eq5}). We extract the embedding of the whole segment of the audio A and audio B and calculate the cosine score in the normal way, then calculate the CMF of the two audios separately, and scale the score using the CMF as the calibration factor.
\begin{equation}
score_{A,B}=CMF_{A}\cdot CMF_{B}\cdot cos(emb_{A}, emb_{B})
\label{eq5}
\end{equation}
When segmenting audio to calculate CMF, there is an overlap of half the segment length so that the segment embedding do not diverge too much. 

For VoxCeleb1-test dataset and VoxSRC23 dev set, the frame length of segment was 400, and overlap was 200. When the audio duration is short, the segment length is appropriately reduced to ensure a certain number of segments. So for the submission of VoxSRC23 test set, the segment frame length was 200 and overlap was 100.

\subsubsection{AS-Norm}
AS-Norm\cite{wang2020dku} was used for all of the models. We selected the original VoxCeleb2 dev dataset without any augmentation. Embeddings of this dataset were averaged speaker-wise, which resulted in 5994 cohorts. Top-400 highest scores are selected to calculate mean and  standard deviation for normalization.

\subsubsection{QMF and fusion}
Quality Measure Functions (QMF) \cite{thienpondt2020idlab} was applied to calibrate the scores, and it greatly enhanced the performance. For QMF, we combined three qualities, speech duration, imposter mean based on AS-Norm, and magnitude of non-normalized embeddings. We selected 30k trials from the original VoxCeleb2-dev as the training set of QMF. Then a Logistic Regression(LR) was trained to serve as our QMF model.

For speech duration, we used duration of enroll and test as quality. In the training data design, audio with duration longer than 5s was considered as long audio. We took  the audio clipped from 2s to 5s as the short audio. Short to short, long to long, long to short, there were 10k pairs for each of the three parts. The ratio of target to nontarget is 1:1.

We fused the single system scores after AS-Norm, and then used QMF to calibrate the fused score.

\section{Training}
We used Pytorch to train our systems. All of our models were trained through two stages. 
In the first stage, we used all VoxCeleb data with speed perturbation, consisting of 17982 speakers. We used 60 GPUs to train ResNet518 with 10 batchsize in each GPU. For other systems, we used 10 GPUs for training, and the batchsize on each GPU was from 20 to 80 due to different model size. The SGD optimizer with a momentum of 0.9 and weight decay of 1e-3 was used, and initial learning rate was 0.08 to train all of our models. 200 frames of each sample in one batch were adopted. We adopted ReduceLROnPlateau scheduler with a frequency of validating every 2,000 iterations, and the patience is 2. The minimum learning rate is 1.0e-6, and the decay factor is 0.1. Furthermore, we used AM-Softmax loss function in first stage, and the margin is 0.2 while scale is 35.

The second stage was large-margin-based fine-tuning stage \cite{thienpondt2020idlab}. Firstly, we removed the speed augmented part from the training set, and Only 5,994 classes were left. Secondly, we changed the frame size from 200 to 600 while increased the margin from 0.2 to 0.5. The AM-Softmax loss was replaced by AAM-Softmax loss. The Inter-TopK penalty was removed to make training stable. Finally, We adopted a smaller finetuning learning rate of 8e-5. The learning rate scheduler was almost the same except the decay factor became 0.5.

\section{Results}
The performance was evaluated using the Equal Error Rate (EER) and the minimum Decision Cost Function (DCF) with parameters $C_{FA}=1$, $C_M=1$, and $P_{target}=0.01$ or $P_{target}=0.05$ for different datasets.

Table \ref{tab:tabel_results} shows the evaluation results of all our single and fusion systems on the VoxCeleb1-test dataset and VoxSRC23 dev set. All the results on the table are the results after using CMF, AS-Norm and QMF. From the results,  we can see that larger models, such as ResNet242, 314 and 518, got better performance. Compared with the result of \cite{zhao2021speakin} in VoxCeleb1-test, Our final result of minDCF has a relative improvement of 12.1\% and 6.8\% on VoxCeleb1-E and VoxCeleb1-H test set. For the system of the same size as  \cite{zhao2021speakin} just like ResNet152, our results also have a relative improvement of 4.8\% and 8.0\%  on VoxCeleb1-E and VoxCeleb1-H, respectively.

\begin{table}[h!]
\centering
\caption{Backends on VoxSRC23-val}
\label{tab:tabel_Ablation}
\setlength{\tabcolsep}{1.5mm}{
\begin{tabular}{lccc}
\toprule
\multicolumn{1}{c}{\multirow{2}{*}{\textbf{Methods}}} & \multicolumn{2}{c}{\textbf{VoxSRC23-val}} \\
\cmidrule(lr){2-3}
\multicolumn{1}{c}{}                                 & \textbf{EER(\%)}   & \textbf{DCF$_{0.05}$}   \\
\midrule
ResNet101                                            & 2.8778             & 0.1516               \\ \midrule
+ AS-Norm                                              & 2.5463             & 0.1348               \\
++ QMF                                               & 2.2655             & 0.1296               \\ \midrule
+ CMF                                                 & 2.6555             & 0.1363               \\
++ AS-Norm                                             & 2.4137             & 0.1259               \\
+++ QMF                                               & \textbf{2.1642}             & \textbf{0.1210}              
\\ \bottomrule
\end{tabular}}
\end{table}

In Table \ref{tab:tabel_Ablation}, ResNet101 without backend was used as baseline and we evaluated the performance of each methods of backend on VoxSRC23-val. Compared with baseline, CMF improved performance from 0.1516 to 0.1363 in minDCF. Applying the AS-Norm and QMF further achieved 2.1642\% EER and 0.1210 minDCF. CMF improved the performance by 6.6\% compared to the result (minDCF=0.1296) of backend without CMF.

\begin{table}[h!]
\centering
\caption{QMF and Segment score on VoxSRC23-test}
\label{tab:tabel_cmf}
\setlength{\tabcolsep}{1.5mm}{
\begin{tabular}{lcccccccc}
\toprule
\multicolumn{1}{c}{\multirow{2}{*}{\textbf{Methods}}} & \multicolumn{2}{c}{\textbf{VoxSRC23-test}} \\
\cmidrule(lr){2-3}
\multicolumn{1}{c}{}                                  & \textbf{EER(\%)}    & \textbf{DCF$_{0.05}$}   \\
\midrule
ResNet242 Segment score w/ 3s                   & 2.1870              & 0.1126               \\
ResNet242 Segment score w/ 2s                   & 2.2780              & 0.1190               \\
ResNet242 CMF w/ 3s                             & 2.1490              & 0.1093               \\
ResNet242 CMF w/ 2s                             & \textbf{2.0990}              & \textbf{0.1080}               \\
\midrule
+QMF                                                  & \textbf{1.760}              & \textbf{0.0993}              
\\ \bottomrule
\end{tabular}}
\end{table}

We compared the performance of CMF and segment score method on VoxSRC23-test. In Table \ref{tab:tabel_cmf}, the results already included AS-Norm. From the result we can see that, when segment score is utilized, decreasing the audio length from 3 seconds to 2 seconds leads to a decrease in system performance. However, when using CMF calibration, compared with segment score, performance improved from 0.1190  to 0.1080 in minDCF with 2 seconds segmentation length. Applying the QMF further achieved 1.760\% EER and 0.0993 minDCF on VoxSRC23-test.

\section{Conclusions}
We presented our system description for Track1 and Track2 of the VoxSRC 2023 challenge in this report. We tried larger model, and got better performance. We proposed a consistency-aware score calibration
method which used Consistency Measure Factor(CMF) scaling score, and brought a huge performance boost in this challenge. We also found only using  standard derivation in pooling layer for ResNets can get better performance. The final result of our system was 0.0855 minDCF and 1.5880\% EER. We achieved the first place in Track 1 and second place in Track 2 of VoxSRC 2023.

\bibliographystyle{IEEEtran}
\bibliography{mybib}

\end{document}